\documentclass{emulateapj}

\usepackage{amsmath}

\def\HI{{{\rm HI}}}
\def\H2{{{\rm H}_2}}

\def\CO{{\rm CO}}
\def\UMW{U_{\rm MW}}
\def\DMW{D_{\rm MW}}
\def\MW{{\rm MW}}

\def\Pext{P_{\rm ext}}

\newcommand{\na}{New Astron.}

\journalinfo{}
\submitted{submitted to APJ}

\shorttitle{The environmental dependence of the pressure -- $\H2$ relation}
\shortauthors{Feldmann, Hernandez, \& Gnedin}

\begin{document}

\title{The relation between mid-plane pressure and molecular hydrogen in galaxies: Environmental dependence}

\author{Robert Feldmann\altaffilmark{1,2},
Jose Hernandez\altaffilmark{3}, and
Nickolay Y. Gnedin\altaffilmark{1,2,4}
}
\altaffiltext{1}{Particle Astrophysics Center, 
Fermi National Accelerator Laboratory, Batavia, IL 60510, USA; feldmann@fnal.gov}
\altaffiltext{2}{Kavli Institute for Cosmological Physics and Enrico
  Fermi Institute, The University of Chicago, Chicago, IL 60637 USA} 
\altaffiltext{3}{Illinois Mathematics and Science Academy, 1500 Sullivan Rd., Aurora, IL 60506, USA} 
\altaffiltext{4}{Department of Astronomy \& Astrophysics, The
  University of Chicago, Chicago, IL 60637 USA} 
\begin{abstract}
Molecular hydrogen ($\H2$) is the primary component of the reservoirs of cold, dense gas that fuel star formation in our galaxy. While the $\H2$ abundance is ultimately regulated by physical processes operating on small scales in the interstellar medium (ISM), observations have revealed a tight correlation between the ratio of molecular to atomic hydrogen in nearby spiral galaxies and the pressure in the mid-plane of their disks. This empirical relation has been used to predict $\H2$ abundances in galaxies with potentially very different ISM conditions, such as metal-deficient galaxies at high redshifts. 
Here, we test the validity of this approach by studying the dependence of the pressure -- $\H2$ relation on environmental parameters of the ISM. To this end, we follow the formation and destruction of $\H2$ explicitly in a suite of hydrodynamical simulations of galaxies with different ISM parameters.
We find that a pressure -- $\H2$ relation arises naturally in our simulations for a variety of dust-to-gas ratios or strengths of the interstellar radiation field in the ISM. Fixing the dust-to-gas ratio and the UV radiation field to values measured in the solar neighborhood results in  fair agreement with the relation observed in nearby galaxies with roughly solar metallicity. However, the parameters (slope and normalization) of the pressure -- $\H2$ relation vary in a systematical way with ISM properties. A particularly strong trend is the decrease of the normalization of the relation with a lowering of the dust-to-gas ratio of the ISM. 
We show that this trend and other properties of the pressure -- $\H2$ relation are natural consequences of the transition from atomic to molecular hydrogen with gas surface density.
\end{abstract}


\section{Introduction}
\label{sect:intro}

The abundance of molecular gas in galaxies is set by the complex interplay of various formation and destruction processes operating in a highly turbulent medium. While many of the individual physical mechanisms are relatively well understood, such as the formation of molecular hydrogen on dust grains, its photo-dissociation by ultraviolet (UV) photons in the Lyman-Werner bands, or the importance of dust and $\H2$ self-shielding, e.g., \cite{1978ApJS...36..595D, 1979ApJS...41..555H, 1988ApJ...332..400S, 1996ApJ...468..269D}, we still lack a realistic and coherent picture that links together the formation of molecular gas, star formation, the turbulent, multi-phase structure of the ISM, and the importance of the various feedback channels due to star formation. The molecular content of galaxies is a key diagnostic that provides insights not only into how galaxies evolve and grow their stellar component, but also helps to constrain the properties of the physical processes (including feedback) that operate in the ISM. In addition, the modeling of the molecular ISM provides a crucial theoretical background for the interpretation of molecular gas surveys, such as those expected in the near future with the Atacama Large Millimeter/sub-millimeter Array (ALMA). 

In addition to analytical and numerical models (\citealt{2006ApJ...645.1024P, 2007ApJ...659.1317G, 2008ApJ...680.1083R, 2008ApJ...689..865K, 2009ApJ...693..216K, 2009ApJ...697...55G, 2010ApJ...721..975O, 2011ApJ...728...88G}), empirical correlations inferred from observations of nearby galaxies are often used to predict the molecular gas abundance. In particular, the correlation between the surface density ratio of molecular and atomic ($\HI$) hydrogen $R_{\rm mol}=\Sigma_\H2/\Sigma_\HI$ and the mid-plane pressure $P_{\rm ext}$ (\citealt{2002ApJ...569..157W, 2004ApJ...612L..29B, 2006ApJ...650..933B}) has been included in semi-analytical models (e.g., \citealt{2009MNRAS.396..141D, 2010MNRAS.409..515F}) and numerical simulations (e.g., \citealt{2010MNRAS.405.1491M}) to estimate $\H2$ mass fractions of galaxies and their star formation rates, or in order to predict the global evolution of the $\H2$ baryon content in the universe \citep{2009ApJ...696L.129O}. Implicit in this approach is the assumption that the empirical correlation continues to hold for galaxies with ISM properties that are potentially different from those found in galaxies in the local universe. It is clearly crucial to test this assumption, either observationally \citep{2010ApJ...722..919F}, or, as is the approach of this paper, with the help of numerical simulations that are based on a firm theoretical modeling of the microphysics of the ISM.

A further complication is the fact that $\H2$ masses and surface densities are typically not directly accessible to observations. The kinetic temperature ($\sim{}10$ K) of the bulk of the molecular hydrogen in galaxies is too low, and the gas sufficiently shielded from UV radiation, to populate the excited levels of the rotational ladder at a significant level \citep{1982ARA&A..20..163S}. Hence, $\H2$ masses are often inferred from the emission of  tracer elements and molecules. In particular, the optically thick emission from the main isotope of carbon-monoxide (CO) serves as a relatively reliable tracer of $\H2$ mass, at least under conditions typical of molecular clouds in the Milky Way \citep{1975ApJ...202...50D, 1986ApJ...309..326D, 1986A&A...154...25B, 1988ApJ...325..389M, 1994ApJ...429..694L, 1996A&A...308L..21S, 1997ApJ...481..205H, 2001ApJ...547..792D, 2007ApJ...663..866D, 2010ApJ...710..133A, 2012arXiv1202.4039T}. However, the conversion factor between $\CO$ luminosity and $\H2$ mass is expect to change systematically with the dust-to-gas ratio and with the strength of the interstellar radiation field \citep{1995ApJ...448L..97W, 1996PASJ...48..275A, 2002A&A...384...33B, 2005A&A...438..855I, 2011MNRAS.412..337G, 2011ApJ...731...25K, 2011MNRAS.412.1686S, 2011ApJ...737...12L, 2012ApJ...746...69G, 2012ApJ...747..124F, 2012MNRAS.421.3127N}. This effect needs to be included in the numerical modeling of empirical relations that are based on CO observations.

In this paper we use numerical simulations of galaxies in a cosmological framework to study the origin of the  $P_{\rm ext} - R_{\rm mol}$ relation and its dependence on the properties of the ISM. Our numerical models compute the local $\HI$ and $\H2$ abundances in the ISM based on a chemical network of well understood formation and destruction processes. In addition, we compute the CO emission expected from our model galaxies to account for variations of the $\CO-\H2$ conversion factor. We show that, under Milky Way like ISM conditions, a $P_{\rm ext} - R_{\rm mol}$ relation similar to that seen in nearby galaxies arises rather naturally. We further show that galaxies with different dust-to-gas ratios and/or radiation fields also follow a $P_{\rm ext} - R_{\rm mol}$ relation, but with changes in the normalization and the slope. We conclude that $\H2$ abundances estimated from a $P_{\rm ext} - R_{\rm mol}$ relation are not robust if these changes are not taken into account.

The outline of the paper is as follows.
In section \ref{sect:methods} we describe the set-up of our numerical approach, discuss details of the data analysis, and introduce the observational data sets that we use to compare with our numerical models. In the subsequent section \ref{sect:results} we present our numerical predictions for the $P_{\rm ext} - R_{\rm mol}$ relation and its dependence on environmental parameters. We also discuss the role of the $\CO$-$\H2$ conversion factor and present a physical model that captures many of the properties of the $P_{\rm ext} - R_{\rm mol}$ relation. We summarize our results and conclude in section \ref{sect:summary}.

\section{Methodology}
\label{sect:methods}

\subsection{Simulations}
\label{sect:sims}

\begin{table*}
\begin{center}
\caption{Details of the numerical simulations and the analyzed snapshots}
\begin{tabular}{lccccc}
\tableline \tableline
label & $\Delta{}x$ & fixed ISM & $D_\MW$ & $U_\MW$ & redshift \\ 
\tableline \noalign{\smallskip}
MW-$d$-$u$  &  65 pc & yes & $d=0.03, 0.1,0.3,1,3$ & $u = 0.1,1,10,100$ & -- \\
MW   &  65 pc & no & -- & -- & $z=3$  \\
MW   &  130 pc & no & -- & -- & $z=1$  \\
\tableline \tableline
\end{tabular}
\label{tab:sims}
\tablecomments{
The first two columns provide the name of the simulation and the peak grid resolution of the analyzed snapshots. The third column specifies whether dust-to-gas ratio $D_\MW$ and the strength of the UV interstellar radiation field $U_\MW$ are determined self-consistently, or whether ``fixed ISM'' conditions are used. In the latter case the range of the imposed $D_\MW$ and $U_\MW$ values is provided in the following two columns. The cosmological redshift of the analyzed snapshots of the self-consistent run is given in the final column.}
\end{center}
\end{table*}

We use the adaptive mesh refinement code ART \citep{1997ApJS..111...73K, 2002ApJ...571..563K} to simulate the formation and evolution of the baryons and dark matter of a Milky Way (MW) sized halo (total mass $\sim{}10^{12}$ $M_\odot$ at $z=0$) in a 6 Mpc h$^{-1}$ cosmological box.  We rely on the standard ``zoom-in'' method of embedding the Lagrangian region of the halo into layers of lower dark matter resolution to reduce the overall computational cost, while capturing the large scale tidal fields correctly \citep{1991ApJ...368..325K, 2001ApJS..137....1B}. The dark matter particle masses are $m_{\rm DM} = 9\times{}10^5$ $M_\odot$ h$^{-1}$ in the highest resolution region and increase by factors of 8 in subsequent lower resolution envelopes. The simulations start from cosmological initial conditions with $\Omega_{\rm m}=0.3$, $\Omega_\Lambda=0.7$, $\Omega_{\rm b}=0.043$, $H_0=70$ km s$^{-1}$ Mpc$^{-1}$, and $\sigma_8=0.9$.

One of our simulations is run fully self-consistently down to $z=1$. In contrast,  ``fixed ISM conditions'' (see below) are imposed on all the other simulations at $z=4$. The latter runs are continued for additional 600 Myr before they are analyzed. At this time the high-resolution Lagrangian region harbors a large disk galaxy sitting in a halo of virial mass $\sim{}4.2\times{}10^{11}$ $M_\odot$ and several lower mass galaxies.

We ``fix ISM conditions'' (see \citealt{2011ApJ...728...88G}) by imposing a spatially uniform dust-to-gas ratio that is independent of the gas metallicity. The gas-to-dust ratio is a crucial parameter that affects the dust shielding and the formation rates of $\H2$. It also enters the CO emission model. We specify the dust-to-gas ratio $D_\MW$ in units of the dust-to-gas ratio in the solar neighborhood. Furthermore, we fix the normalization of the radiation field at 1000 {\AA} to $J_0$ = 10$^6$ photons cm$^{-2}$ s$^{-1}$ sr$^{-1}$ eV$^{-1}$, a value typical for the solar neighborhood in the Milky Way \citep{1978ApJS...36..595D, 1983A&A...128..212M}. We use the notation $\UMW=1$, where $\UMW$ is the intensity of the radiation field at 1000 {\AA} in units of $J_0$. We stress that only the normalization of the radiation field is fixed. The shape of the radiation spectrum is not modified.

All simulations include a photo-chemical network that follows the formation and destruction of molecular hydrogen in addition to the five major atomic and ionic species of hydrogen and helium. The simulations also include metal enrichment from supernova (type Ia and type II), but no thermal energy injection, optically thin radiative cooling by hydrogen (including $\H2$), helium, and metal lines, and 3D radiative transfer of ionizing and non-ionizing UV radiation from stellar sources in the Optically Thin Variable Eddington Tensor (OTVET) approximation \citep{2001NewA....6..437G}. The details of the implementation can be found in \cite{2009ApJ...697...55G} and \cite{2011ApJ...728...88G}.

We compute the emission arising from the $J=1\rightarrow{}0$ rotational transition of the $^{12}\mathrm{C}^{16}\mathrm{O}$ isotope in a post-processing step as described in \cite{2012ApJ...747..124F}. In brief, the CO abundance in each simulation grid cell is computed based on the results of a suite of small scale magneto-hydrodynamical ISM simulations \citep{2011MNRAS.412..337G}. The emission is then computed using the escape probability formalism and assuming a virial scaling of the $\CO$ line width. The contributions from these individual, $\sim{}65-130$ pc sized resolution elements are then combined in the optically thin limit to derive the CO emission from larger regions. The $\CO$ brightness temperature is a free parameter of the model. We adopt an excitation temperature of 10 K (approximately the kinetic temperature of molecular clouds in the Milky Way) and a corresponding brightness temperature against the CMB background of 6.65 K. 

An overview of the set of the analyzed simulation snapshots is given in table \ref{tab:sims}.

\subsection{Measuring $P_{\rm ext}$ and $R_{\rm mol}$}

The ISM pressure in the mid-plane of a disk galaxy is not easily accessible to observations. In order to arrive at an estimate, disk galaxies are often modeled as self-gravitating two-component disks, consisting of a turbulent gas layer and a stellar disk, in hydrostatic equilibrium \citep{1989ApJ...338..178E}. If these assumptions are made, the mid-plane pressure can be expressed in terms of the gas and stellar surface densities ($\Sigma_g$, $\Sigma_*$) and the velocity dispersions of the gas and stellar disk ($\sigma_g$, $\sigma_*$):
\begin{equation}
\label{eq:Pext1}
P_{\rm ext} \simeq{} \frac{\pi}{2}G\Sigma_g\left(\Sigma_g + \Sigma_* \frac{\sigma_g}{\sigma_*}\right).
\end{equation}
The velocity dispersions can be computed for virialized disks and equation (\ref{eq:Pext1}) can be rewritten using the scale heights $h_g$ and $h_*$ of the gas and stellar disks \citep{2004ApJ...612L..29B, 2006ApJ...650..933B}:
\begin{equation}
\label{eq:Pext2}
\Pext \simeq{} 0.84 G^{0.5} \Sigma_g \sigma_g \left[ \left(\frac{\Sigma_*}{h_*}\right)^{0.5} + \left(\frac{\pi}{4}\frac{\Sigma_g}{h_g}\right)^{0.5}\right].
\end{equation}
In this paper we follow Blitz \& Rosolowski and adopt equation (\ref{eq:Pext2}), with $\Sigma_g$ and $\Sigma_*$ as the independent variables. In addition, we adopt their choice $\sigma_g=8$ km s$^{-1}$. We also decided to fix the values of the gaseous and stellar scale heights to $h_{\rm g}=100$ pc and $h_*=300$ pc, typical of disk galaxies in the local universe. This is done, because in our simulations the disk scale heights are, at best, only marginally resolved. However, since the scale heights enter 
(\ref{eq:Pext2}) in form of a square root, none of our results change significantly if the scale heights are varied within reasonably bounds.

$\H2$ masses and surface densities are often derived from $\CO$ observations using the galactic $\CO$-$\H2$ conversion factor. Its numerical value has been determined to within a factor of two by a number of independent techniques (e.g., \citealt{1987ApJ...319..730S, 1996A&A...308L..21S, 2001ApJ...547..792D}) and a commonly adopted value is $X_\CO=N_\H2/W_\CO\sim{}2\times{}10^{20}$ cm$^{-2}$ K$^{-1}$ km$^{-1}$ s (without He), or, equivalently, $\alpha_\CO = M_\H2 / L_\CO \sim{} 4.4$ $M_\odot$ pc$^{-2}$ K$^{-1}$ km$^{-1}$ s (including He). 

In order to assess the importance of potential conversion factor variations, we compute in two different ways the $\H2$ surface density that enters the neutral gas surface density $\Sigma_g=\Sigma_\H2+\Sigma_\HI$ (including He) and the surface density ratio 
\begin{equation}
\label{eq:Rmol}
R_{\rm mol}=\frac{\Sigma_\H2}{\Sigma_\HI}.
\end{equation}
The first method takes the $\H2$ surface density directly as predicted by the chemical network in the simulation. Alternatively, we use the $\CO$ emission predicted by our $\CO$ model (see section \ref{sect:sims}) and convert it into an $\H2$ surface density using the galactic conversion factor. In general, this has three effects if the actual conversion factor differs from the adopted galactic value: (1) it changes the minimum $\H2$ surface density that is detectable, (2) it changes the estimate of $P_{\rm ext}$ via the change in $\Sigma_g$, and (3) it changes $R_{\rm mol}$. 

\begin{figure}
\begin{tabular}{c}
\includegraphics[width=80mm]{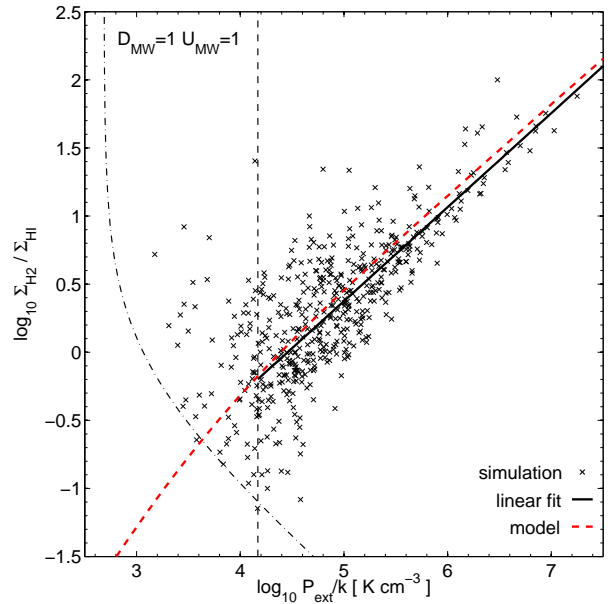}
\end{tabular}
\caption{The $P_{\rm ext} - R_{\rm mol}$ relation as predicted by simulations that follow the formation and destruction of $\H2$ explicitly. The panels show the results for a MW-like dust-to-gas ratio ($D_\MW=1$) and for a MW-like UV radiation field ($\UMW=1$). The black crosses show the pressure inferred from equation (\ref{eq:Pext2}) and the ratio of the $\H2$ surface mass density (inferred from $\CO$ emission with a galactic $\CO$-$\H2$ conversion factor) to $\HI$ surface mass density for 1 kpc$^2$ patches of the ISM above the $\CO$ intensity detection limit of $0.2$ K km s$^{-1}$.
The black solid line is the result of an ordinary linear regression of $\log_{10} R_{\rm mol}$ given $\log_{10} P_{\rm ext}$, in the range between the vertical dashed line (the truncation limit) and $\log_{10}P_{\rm ext}/(k_{\rm B}\,{\rm K}\,{\rm cm}^{-3})=7.5$. The fitted slope is 0.69. The truncation limit is determined as the pressure that corresponds to the lowest $R_{\rm mol}$ value in the sample of kpc$^2$ patches with a $\CO$ emission above the $\CO$ detection limit. The black dot-dashed line separates the regions of data lying below (to the left) or above (to the right) the $\CO$ detection limit. The red dashed lines show the scaling predicted by a simple model (see section \ref{sect:model}).}
\label{fig:Fit}
\end{figure}

\subsection{Fitting the $P_{\rm ext} - R_{\rm mol}$ relation}

In our set of simulations we measure the $P_{\rm ext} - R_{\rm mol}$ relation on kpc$^{2}$ patches (the line of sight depth is 1 kpc) of the ISM. In order to characterize the main properties of the relation, we perform a linear regression in log-log space, i.e., we fit for the normalization A and the slope S:
\[
\log_{10} R_{\rm mol} = A + S\,[\log_{10} \frac{P_{\rm ext}}{k_{\rm B}\,{\rm K}\,{\rm cm}^{-3}} - 5.5].
\]

We follow \cite{2006ApJ...650..933B} in including in our fit only those data points that have $P_{\rm ext}$ larger than a truncation limit to avoid biasing the regression. Including data point at low pressure values can potentially bias the slope low, because low $R_{\rm mol}$ values are preferentially excluded due to the finite $\CO$ detection limit, but it can also bias it high, since at $R_{\rm mol}<1$ the $P_{\rm ext} - R_{\rm mol}$ relation steepens due to the relatively rapid transition between $\HI$ and $\H2$, see section \ref{sect:model}.

The pressure of the data point that is above the detection limit and has the lowest $R_{\rm mol}$ is taken as the truncation limit. For instance, in the fixed ISM simulation with $\UMW=1$ and $\DMW=1$ the truncation limit is $\sim{}1.5\times{}10^4\, k_{\rm B}\,{\rm K}\,{\rm cm}^{-3}$, while in the simulation with $\UMW=100$ and $\DMW=1$ the limit is $\sim{}9\times{}10^4\, k_{\rm B}\,{\rm K}\,{\rm cm}^{-3}$.

\subsection{Observational data}
\label{sect:obsdata}

\cite{2006ApJ...650..933B} tabulate the results of fits to the $P_{\rm ext} - R_{\rm mol}$ relation in their table 2. We take the slope and its error bar directly from their table and convert the stated normalization $P_0$ to our pivot point of $\log_{10}P_{\rm ext}/(k_{\rm B}\,{\rm K}\,{\rm cm}^{-3})=5.5$. We exclude the galaxies NGC 598 and NGC 4414 for which we lack reliable oxygen abundances. We also exclude their Milky Way measurements, because their Fig. 3 shows that most of the data falls into the  $\HI$-$\H2$ transition regime where a single power-law is not a good fit to the data.

\cite{2008AJ....136.2782L} provide in their table 7 radial profiles of the $\HI$, $\H2$ (based on CO observations), and stellar mass surface densities for a sample of nearby galaxies. For each galaxy we compute in each radial bin $P_{\rm ext}$ and $R_{\rm mol}$ according to (\ref{eq:Pext2}) and (\ref{eq:Rmol}) and perform a linear regression in the same way as for our simulated galaxies. We exclude three of the galaxies (NGC2841, NGC3198, and NGC3351) with available $\H2$ data from the fitting, either because 
there are not sufficiently many radial bins above the truncation limit to perform a reliable fit or because a power-law correlation is not apparent in the radial data. 

We note that the binning of the data in radial bins with areas $>$kpc$^2$ potentially biases the fit results. The reason is that the average gas surface density decreases with increasing spatial scale due to the inclusion of regions that fall below the detection limit at higher resolution. Hence, the estimate for $P_{\rm ext}$ decreases and, assuming that $R_{\rm mol}$ is less affected than $P_{\rm ext}$, the normalization of the $P_{\rm ext} - R_{\rm mol}$ relation increases with scale. We therefore expect the \cite{2008AJ....136.2782L} data to be shifted upward w.r.t. the \cite{2006ApJ...650..933B} observations. Also, since the area of the radial bins varies, the measurement of the slope could in principle be affected. However, we find that the derived slopes are similar to those found in \cite{2006ApJ...650..933B} and, hence, conclude that this latter bias cannot be very strong.

In order to study the $P_{\rm ext} - R_{\rm mol}$ relation as function of dust-to-gas ratio we supplement the galaxies in both data sets with gas-phase oxygen abundances \citep{2010ApJS..190..233M} as a proxy for the dust-to-gas ratio. As discussed by the authors the absolute calibration of the oxygen abundance is rather uncertain (up to $\sim{}0.6$ dex) and depends crucially on the adopted methodology. We therefore follow the suggestion of the authors and use the average of the characteristic abundances (their table 9) derived from a theoretical \citep{2004ApJ...617..240K} and from an empirical  \citep{2005ApJ...631..231P} calibration method. Half the difference between the two methods is used as a measure of the systematic error which we then combine with the provided error for each of the individual methods to obtain a total error estimate.

\section{The origin of the $P_{\rm ext} - R_{\rm mol}$ relation}
\label{sect:results}

\subsection{Predictions of the numerical simulations}
\label{sect:SimPred}

In Fig.~\ref{fig:Fit} we show $P_{\rm ext}$, using equation (\ref{eq:Pext2}), and $R_{\rm mol}$, using equation (\ref{eq:Rmol}), for kpc$^2$ patches of the ISM of a simulated galaxy with MW-like ISM conditions. The $\H2$ surface density that enters $P_{\rm ext}$ and  $R_{\rm mol}$ is calculated using the $\CO$ emission from each patch. The results remain essentially unchanged if we use the actual $\H2$ surface densities computed in the simulation, because, for a MW-like dust-to-gas ratio, the $\CO$-$\H2$ conversion factor is close to the canonical galactic value over a wide range of $\H2$ surface densities (see Fig.~10 in \citealt{2012ApJ...747..124F}).

The figure demonstrates that a power-law relation between $P_{\rm ext}$ and $R_{\rm mol}$ arises naturally in simulations that follow the microphysics of $\H2$ formation and destruction in the ISM. The figure also includes the prediction of a simple model which approximates the simulation results rather well. We will introduce and discuss this model in section \ref{sect:model}.

\begin{figure*}
\begin{tabular}{cc}
\includegraphics[width=80mm]{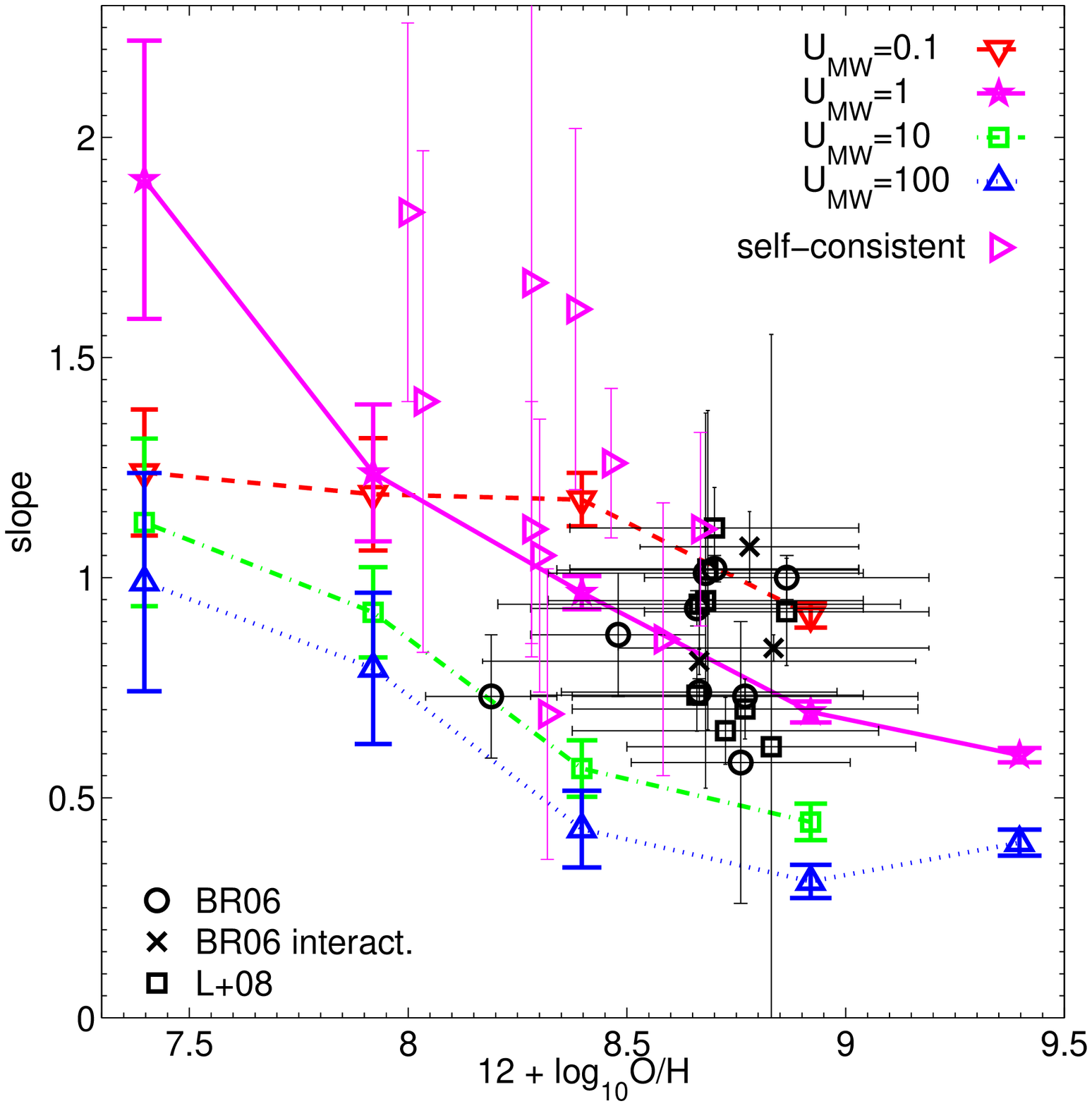} &
\includegraphics[width=80mm]{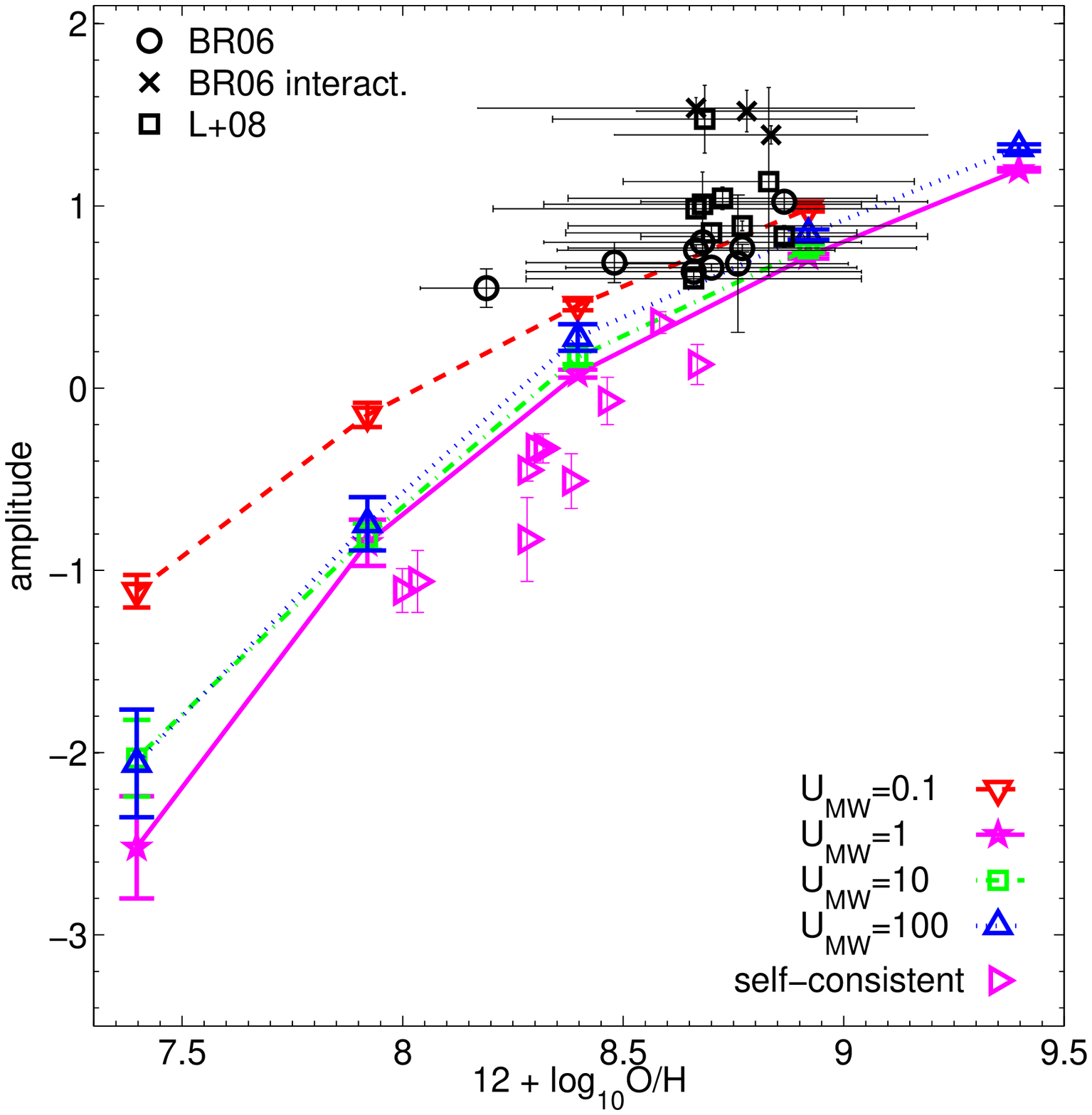} \\
\end{tabular}
\caption{The slope (left panel) and normalization (right panel) of the $\log_{10} P_{\rm ext} - \log_{10} R_{\rm mol}$ relation as function of metallicity. Black circles, x-s, and squares show observational data from \cite{2006ApJ...650..933B} and \cite{2008AJ....136.2782L}, see legend. 
The results from the ``fixed ISM'' simulations are shown as lines with error bars for four different values of the UV radiation field (see legend). The magenta left-ward pointing triangles show the combined $z=3$ and $z=1$ results of the fully self-consistent run, see table~\ref{tab:SelfCons}. The x-axis of the observational data refers to gas-phase oxygen abundances derived from nebular emission lines originating in HII regions \citep{2010ApJS..190..233M}.  In contrast, the x-axis of the simulation data refers to the dust-to-gas ratio which we convert into an oxygen abundance via $12+\log_{10} {\rm O}/{\rm H} =  \log_{10} D_\MW + 8.92$. Note this assumes that oxygen abundance and dust-to-gas ratio scale proportional to each other down to low metallicities. 
Error bars of the slope and normalization only include fit errors and are estimated via bootstrapping. The error bars in the gas-phase oxygen abundance are dominated by systematic uncertainties (see section \ref{sect:obsdata}). Slopes and normalizations for simulated galaxies with MW-like ISM conditions are in fair agreement with observational estimates for a sample of nearby (many of them MW-like) galaxies (see text for a discussion of the outliers). Interestingly, the simulations predict significant changes in the slope and normalization with changing ISM parameters. In particular, a decrease in the dust-to-gas ratio is predicted to reduce the normalization of the $\log_{10} P_{\rm ext} - \log_{10} R_{\rm mol}$ relation.}
\label{fig:FitObs}
\end{figure*}

In Fig.~\ref{fig:FitObs} we compare the slope and normalization of the $\log_{10} P_{\rm ext} - \log_{10} R_{\rm mol}$ relation as obtained from our numerical experiments with the observations by \cite{2006ApJ...650..933B} and \cite{2008AJ....136.2782L}. Overall simulations and observations are in reasonable agreement although there are a few interesting outliers. 

Notably is the higher normalization of three galaxies in the \cite{2006ApJ...650..933B} sample. As pointed out by the authors these three galaxies are in an interaction stage and hence the choice $\sigma_g=8$ km s$^{-1}$ may underestimate the actual gas velocity dispersion. If this is indeed so, then $P_{\rm ext}$ is underestimated in these galaxies and, hence, the normalization is overestimated. Interestingly, one of the galaxies in the \cite{2008AJ....136.2782L} sample, NGC 628 also has a normalization significantly above the numerical predictions. However, the values of $P_{\rm ext}$ in the radial bins of this galaxy do not exceed $\log_{10}P_{\rm ext}/(k_{\rm B}\,{\rm K}\,{\rm cm}^{-3})=5.2$ and, hence, the normalization at the pivot point is based on an extrapolation. 

Focusing on the predictions of the ``fixed ISM'' simulations, we find that both slope and normalization change with dust-to-gas ratio and the UV radiation field in the ISM. Consequently, the relation between $P_{\rm ext}$ and $R_{\rm mol}$ is not universal. In other words, if one wants to predict the $\H2$ abundance based on the mid-plane pressure in galaxies one needs to take the systematic changes of the slope and normalization as function of dust-to-gas ratio and UV radiation field into account.
The changes are in particular
\begin{itemize}
\item a decrease in the normalization with decreasing dust-to-gas ratio,
\item an increase in the slope with decreasing dust-to-gas ratio, and
\item a decrease in the slope with increasing UV radiation field.
\end{itemize}
We discuss the origin of these trends in section \ref{sect:model}.

\begin{table}
\begin{center}
\caption{Resolution study}
\begin{tabular}{|r|c|r|r|}
\tableline \tableline
$\Delta{}x$ & method & slope & amplitude \\ \tableline
130 pc & $\CO$ & $0.73\pm{}0.05$ & $0.77\pm{}0.02$ \\
65 pc & $\CO$ & $0.69\pm{}0.02$ & $0.72\pm{}0.01$ \\
32 pc & $\CO$ & $0.69\pm{}0.03$ & $0.75\pm{}0.02$ \\
130 pc & $\H2$ & $0.71\pm{}0.03$ & $0.82\pm{}0.02$ \\
65 pc & $\H2$ & $0.67\pm{}0.02$ & $0.79\pm{}0.01$ \\
32 pc & $\H2$ & $0.69\pm{}0.03$ & $0.80\pm{}0.02$ \\
\tableline \tableline
\end{tabular}
\label{tab:ResStudy}
\tablecomments{Resolution study of a simulation with constrained ISM conditions ($\UMW=1$, $\DMW=1$). The first column states grid size of the highest resolution elements in each simulation. The second column denotes whether the $\H2$ surface densities are taken directly from the simulation (``$\H2$''), or whether they are estimated based on $\CO$ intensities calculated with the $\CO$ model described in section \ref{sect:sims}. The third and fourth column give the slope and amplitude of the $\log_{10} P_{\rm ext} - \log_{10} R_{\rm mol}$ relation, respectively, and also show 1-$\sigma$ bootstrap errors of the linear fit. The fit results are converged for $\Delta{}x\lesssim{}130$ pc.}
\end{center}
\end{table}

In order to ensure that our predictions are not affected by numerical resolution we have re-simulated the run with MW-like ISM condition ($\DMW=1$, $\UMW=1$) both at a two times lower and a two times higher spatial resolution. Table~\ref{tab:ResStudy} shows that our numerical predictions are converged.

\begin{table*}
\begin{center}
\caption{Fits to the pressure -- $\H2$ relation for simulated galaxies}
\begin{tabular}{|c|c|c|c|c|c|c|c|}
\tableline \tableline
Galaxy-ID & $z$ & $M_*$ & $\langle{}\DMW\rangle{}$ & $\langle{}\UMW\rangle{}$ &  slope &  amplitude \\ 
                   &         &  ( $10^{10} M_\odot$ ) &                                               &                     &                            &       \\ \tableline
1 & 3 & 4.2 & 0.46 & 3.2  & $0.86\pm{}0.31$ & $0.36\pm{}0.06$   \\
2 & 3 & 0.92 & 0.29 & 0.73  & $1.61\pm{}0.41$  & $-0.51\pm{}0.15$  \\
3 & 3 & 0.33 & 0.23 & 0.36 & $1.67\pm{}0.82$  & $-0.83\pm{}0.23$   \\
4 & 3 & 0.24 & 0.12 & 0.30 & $1.83\pm{}0.43$   & $-1.11\pm{}0.12$ \\
5 & 3 & 0.22 & 0.13 & 0.28 & $1.40\pm{}0.57$  & $-1.06\pm{}0.17$  \\
6 & 1 & 11.3 & 0.35 & 2.4 & $1.26\pm{}0.17$  & $-0.07\pm{}0.13$  \\
7 & 1 & 4.1 & 0.56 & 1.2 & $1.11\pm{}0.22$ & $0.13\pm{}0.11$ \\ 
8 & 1 & 0.64 & 0.25 & 0.15 & $0.69\pm{}0.33$   & $-0.33\pm{}0.08$ \\
9 & 1 & 0.44 & 0.23 & 0.18 & $1.11\pm{}0.29$  & $-0.45\pm{}0.06$  \\
10 & 1 & 0.28 & 0.24 & 0.07 & $1.05\pm{}0.31$  & $-0.33\pm{}0.07$  \\
\tableline \tableline
\end{tabular}
\label{tab:SelfCons}
\tablecomments{Properties of self-consistently simulated galaxies at two different redshifts. The first column provided the identifier of the simulated galaxy. Further column contain: (2) redshift, (3) the stellar mass within 10 kpc from the center of the galaxy, (4) the mass-weighted average dust-to-gas ratio within 10 kpc (relative to the Milky Way) , (5) the volume-weighted average UV radiation field within 10 kpc (relative to the Milky Way),  (6) the slope of the $\log_{10} P_{\rm ext} - \log_{10} R_{\rm mol}$ relation, and (7) the amplitude of the relation. The spatial resolution is $65$ pc at $z=3$ and $130$ pc at $z=1$.}
\end{center}
\end{table*}

Fig.~\ref{fig:FitObs} also includes the fit results for galaxies taken from the $z=1$ and $z=3$ snapshots of our fully self-consistent simulation. Although some differences can be seen (e.g., the slightly lower predictions for the normalization), they follow more or less the predictions for an ISM with a MW-like UV radiation field. This is not surprising since the (volume-weighted) average $\UMW$ value in these galaxies is typically in the range 0.3-3. Table~\ref{tab:SelfCons} summarizes the relevant global properties of the simulated galaxies in the self-consistent run. 

\subsection{The role of the $\CO$-$\H2$ conversion factor}

It has been demonstrated both observationally, e.g., \cite{1995ApJ...448L..97W, 2011ApJ...737...12L}, and theoretically, e.g., \cite{2011ApJ...731...25K, 2012ApJ...747..124F}, that the $\CO$-$\H2$ conversion factor increases with decreasing metallicity or dust-to-gas ratio of a galaxy. Consequently, the $\H2$ surface mass density derived from $\CO$ emission will be biased if a constant conversion factor, e.g., the canonical galactic value, is used for a galaxy with a dust-to-gas ratio or metallicity very different from that of the Milky-Way. If so, the estimates for $P_{\rm ext}$ and $R_{\rm mol}$ will be affected and, therefore, the slope and normalization of the $\log_{10} P_{\rm ext} - \log_{10} R_{\rm mol}$ relation. In fact, the fit results that are shown in Fig.~\ref{fig:FitObs} use a galactic conversion factor to convert the $\CO$ emission into an $\H2$ surface density. This was done to allow for a fair comparison between our simulations and the observations by \cite{2006ApJ...650..933B} and \cite{2008AJ....136.2782L}. However, one may wonder whether, and to which extent, some of the trends with dust-to-gas ratio seen in Fig.~\ref{fig:FitObs} are an artifact of using an incorrect $\CO$-$\H2$ conversion factor.

\begin{figure*}
\begin{tabular}{cc}
\includegraphics[width=80mm]{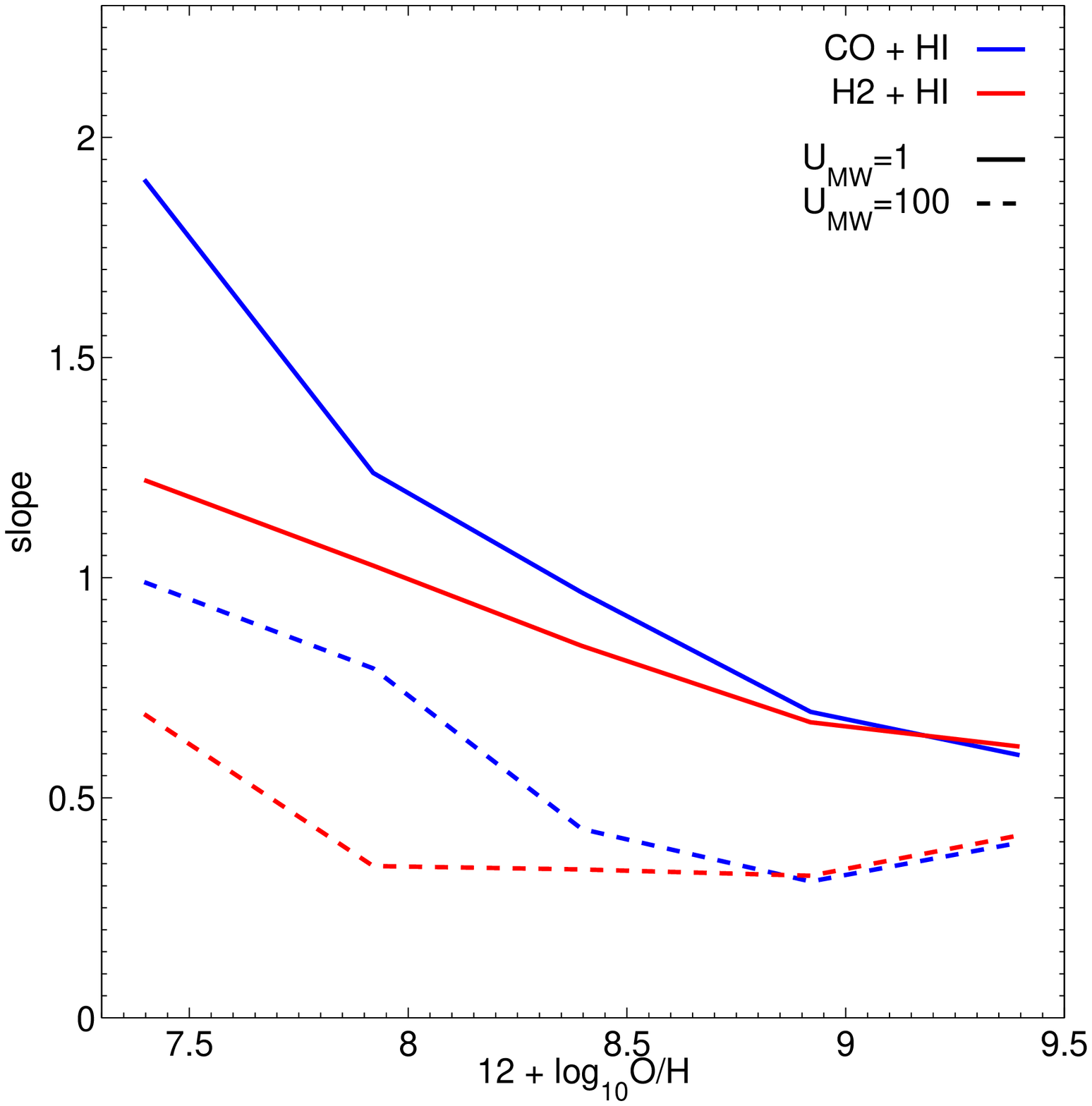} &
\includegraphics[width=80mm]{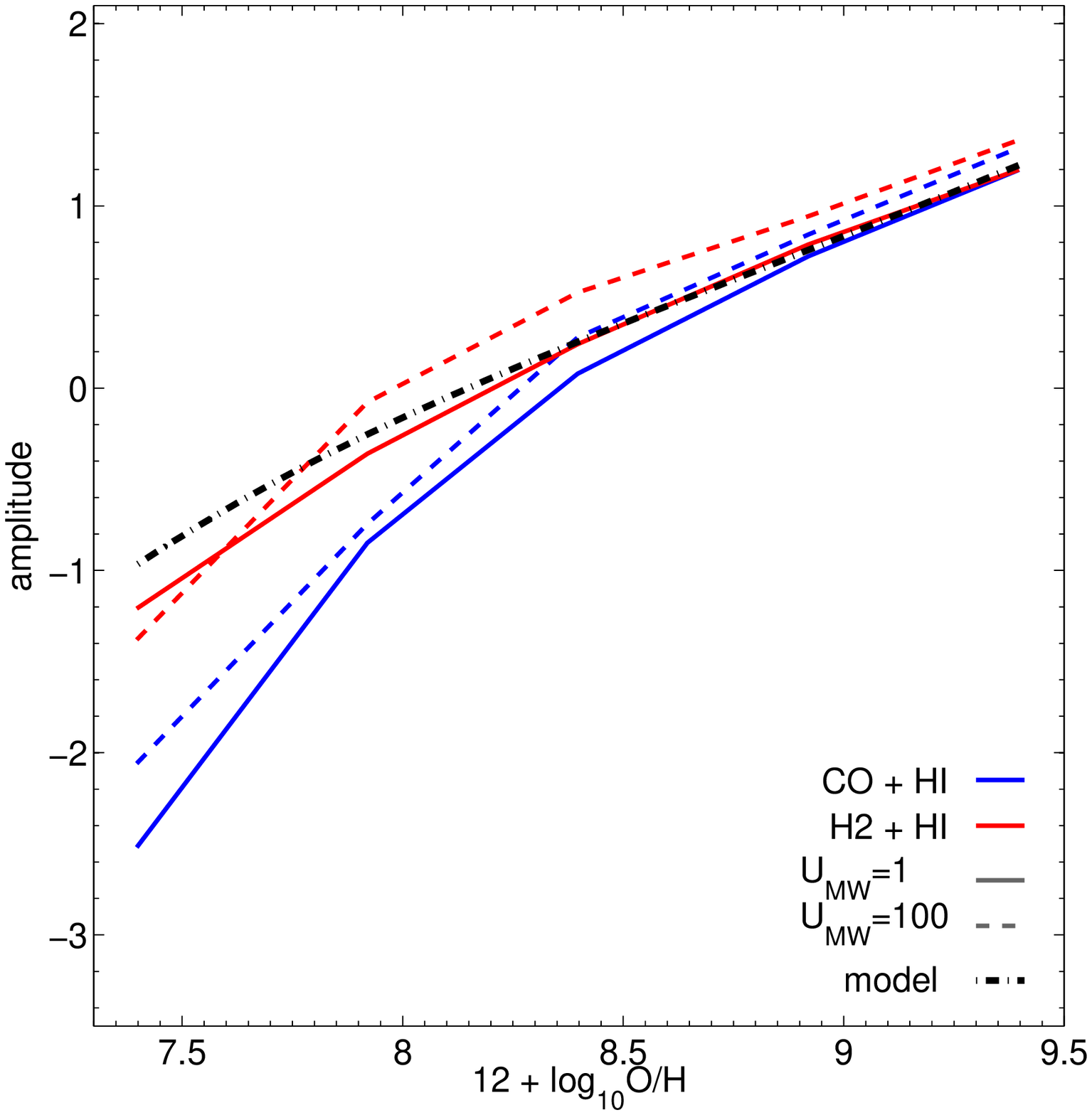} \\
\end{tabular}
\caption{The impact of the $\CO$-$\H2$ conversion factor on the slope (left panel) and normalization (right panel) of the $\log_{10} P_{\rm ext} - \log_{10} R_{\rm mol}$ relation. Shown are the fit results based on simulations with a fixed dust-to-gas ratio $\DMW$ converted to oxygen abundance via $12+\log_{10} {\rm O}/{\rm H} =  \log_{10} D_\MW + 8.92$ and with a UV radiation field $\UMW=1$ (solid lines) or $\UMW=100$ (dashed lines). Biases in the $\H2$ surface density estimates can affect the fit since both $P_{\rm ext}$ and $R_{\rm mol}$  depend on the $\H2$ surface density. Blue lines show the fit results if the $\H2$ surface density estimate is based on the $\CO$ emission assuming a galactic conversion factor, while red lines use the $\H2$ surface density taken directly from the simulation. The black dot-dashed line shows the prediction of the simple model described in section \ref{sect:model}. While some of the trends are aggravated by the dependence of the conversion factor on the dust-to-gas ratio, most of the trends remain qualitatively unchanged if the true $\H2$ surface density is used and conversion factor effects are eliminated.}
\label{fig:FitTheo}
\end{figure*}

To address this question we compare in Fig.~\ref{fig:FitTheo} the slopes and normalizations that we obtain using either $\CO$-based $\H2$ surface densities (and assuming a galactic conversion factor) or by taking the $\H2$ surface densities directly from the simulations. The figure shows that the decrease in the normalization of the $P_{\rm ext}- R_{\rm mol}$ relation with decreasing metallicity is also present if the true $\H2$ surface densities are used, showing that it is not primarily a result of the metallicity dependence of the $\CO$-$\H2$ conversion factor. The latter does, however, aggravate the decrease in normalization at low dust-to-gas ratios. The conversion factor is also not responsible for the lowering of the slopes with increased UV radiation field. It does, however, contribute to the increase in slope with decreasing dust-to-gas ratio.

\subsection{Modeling the $P_{\rm ext} - R_{\rm mol}$ relation}
\label{sect:model}

The results presented in section \ref{sect:SimPred} show that a relation between $P_{\rm ext}$  and $R_{\rm mol}$ arises naturally if $\H2$ formation and destruction processes in the ISM are followed in a self-consistent manner. We have also demonstrated that the numerical predictions are in agreement with observational data for nearby, MW-like galaxies and, in addition, made specific prediction for the scaling of the slope and normalization of the relation with dust-to-gas ratio and UV radiation field of a given galaxy. In this section, we will discuss the origin of these trends and show that many of them can be understood as a consequence of the $\HI$ to $\H2$ transition.

Many observational studies fit the $P_{\rm ext} - R_{\rm mol}$ relation in the $\H2$ dominated regime, i.e., for data with $R_{\rm mol}>1$. In this regime the atomic hydrogen is close to its saturation limit (e.g., \citealt{2011ApJ...728...88G}) and, hence, to first order, $R_{\rm mol} \propto{} \Sigma_g$. Let us now assume that either (case A) $\Sigma_*/h_* \ll{} \Sigma_g/h_g$ (this, for instance, holds in a gas-dominated galaxy), or that (case B) $\Sigma_*\propto{}\Sigma_g^\beta$. Combining equations (\ref{eq:Pext2}) and (\ref{eq:Rmol}) we then obtain $R_{\rm mol}\propto{}P_{\rm ext}^{2/3}$ in case A and $R_{\rm mol}\propto{}P_{\rm ext}^{2/(2+\beta)}$ in case B. 

This simple analysis therefore predicts that $R_{\rm mol}$ is correlated with $P_{\rm ext}$ and that the slope of the $\log_{10} P_{\rm ext} - \log_{10} R_{\rm mol}$ relation in the $\H2$ dominated regime is $\sim{}0.67$ or $2/(2+\beta)$ depending on whether the density in the mid-plane is dominated by the ISM or the stellar component. In fact, a slope of the order of $\sim{}0.7$ is indeed what we find in our numerical simulations for the case $\UMW\sim{}1$, see Fig.~\ref{fig:Fit} and~\ref{fig:FitObs}. The observations by \cite{2006ApJ...650..933B} and \cite{2008AJ....136.2782L} are largely consistent with this slope although with a substantial scatter that includes galaxies with significantly larger slopes.

We can quantify these arguments further by using a simple model that is based on the numerical study of the $\HI$ to $\H2$ transition by \cite{2011ApJ...728...88G}. It is therefore, ultimately, based on the small scale physics of $\H2$ formation and destruction in the ISM. It captures the basic mechanisms behind the $P_{\rm ext} - R_{\rm mol}$ relation, as shown in Fig.~\ref{fig:Fit} and Fig.~\ref{fig:FitTheo}. It works as follows.

\begin{figure*}
\begin{tabular}{cc}
\includegraphics[width=80mm]{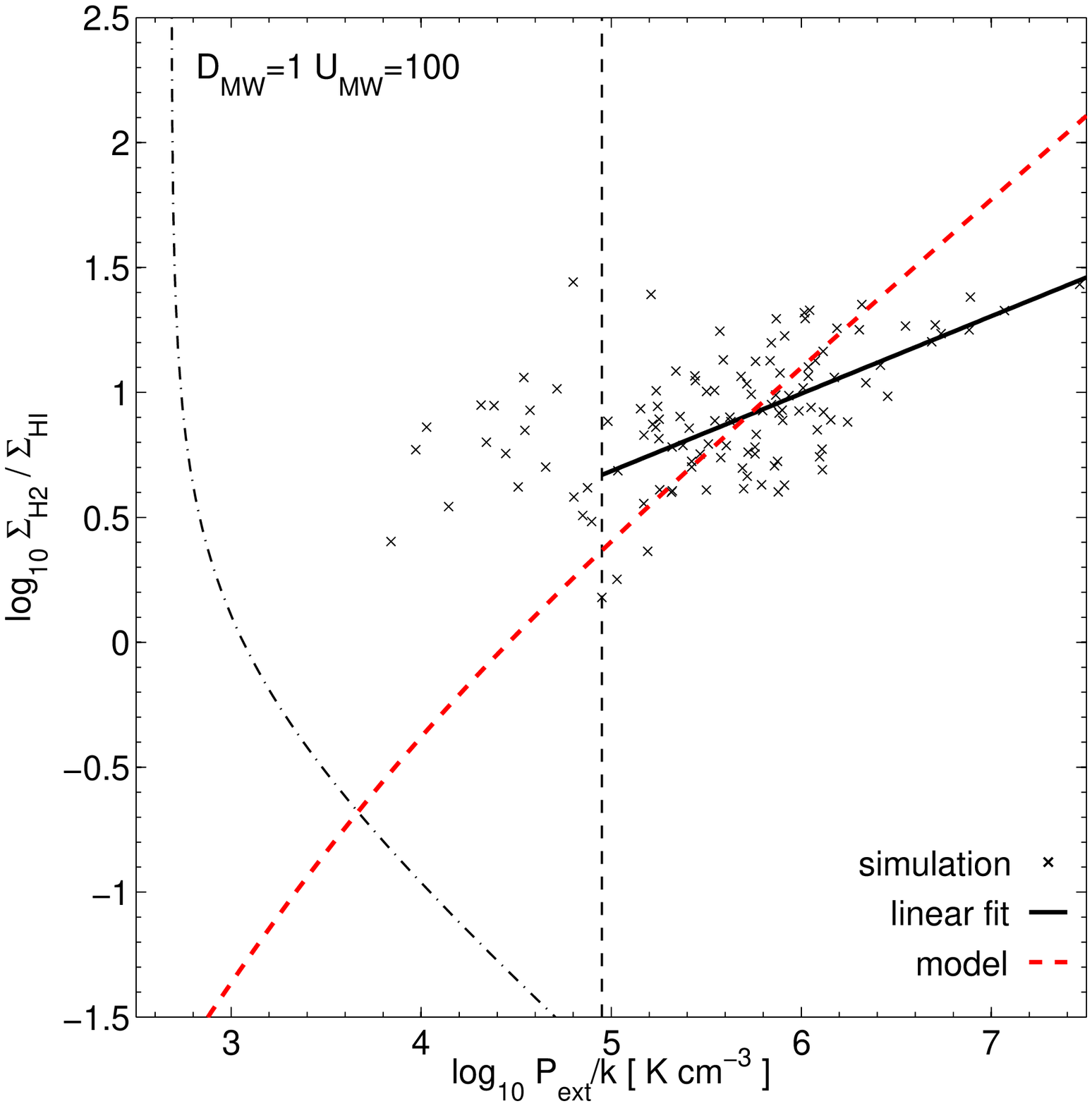} &
\includegraphics[width=80mm]{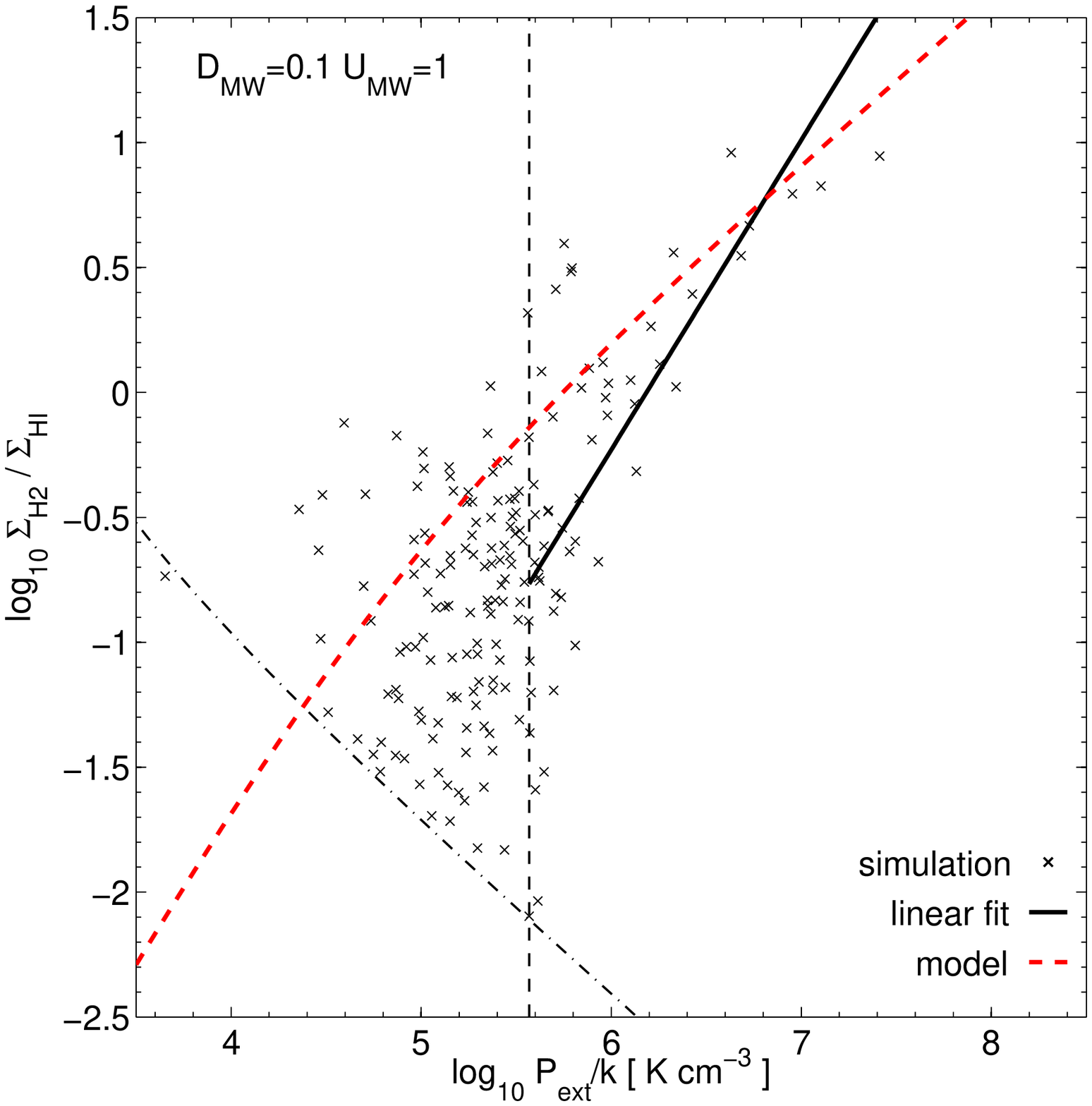} \\
\end{tabular}
\caption{Same as Fig.~\ref{fig:Fit}, but for a large UV radiation field ($\UMW=100$, left panel) and a low dust-to-gas ratio ($\DMW=0.1$, right panel), respectively. The simple model (see text) fails at reproducing the slopes of the $\log_{10} P_{\rm ext} - \log_{10} R_{\rm mol}$ relation recovered via a linear regression for the reasons outlined in the text, but it does reproduce the normalization of the data points above the truncation limit.}
\label{fig:Fit2}
\end{figure*}

The $\H2$ and $\HI$ surface densities are computed as
\begin{equation}
\label{eq:ModelSigmas}
\Sigma_\H2 = \frac{\Sigma_g}{(1+\Sigma^\infty_\HI/(2\Sigma_g))^2}\,\,\,{\rm and} \,\,\,\Sigma_\HI = \Sigma_g- \Sigma_\H2,
\end{equation}
where $\Sigma_\HI^\infty$ is the $\HI$ saturation limit (see below), which is a function of dust-to-gas ratio and interstellar UV field. With $\Sigma_\HI$ and $\Sigma_\H2$ given, $R_{\rm mol}$ can be computed for any given $\Sigma_g$. In order to compute $P_{\rm ext}$ via equation (\ref{eq:Pext2}) an estimate of the stellar surface density $\Sigma_*$ is required. The lines shown in Fig.~\ref{fig:Fit}, \ref{fig:FitTheo}, and \ref{fig:Fit2} assume that $\Sigma_*=\Sigma_g$. We note that most of the galaxies in the \cite{2008AJ....136.2782L} sample
show a scaling $\Sigma_*\propto{}\Sigma_g^\beta$ with $\beta\sim{}1-2$ in the $\H2$ dominated regime.

The $\HI$ saturation limit is given by the fitting formula
\begin{equation}
\Sigma_\HI^\infty = 20\,M_\odot\,{\rm pc}^{-2}\,\frac{\Lambda^{4/7}}{\DMW}\frac{1}{(1+\DMW\,\UMW)^{0.25}}.
\end{equation}
In the $\H2$ dominated regime this formula is significantly more accurate than the corresponding expression (their equation (14)) provided by \cite{2011ApJ...728...88G}. The $\Lambda$ parameter depends on $\UMW$ and $\DMW$ as follows \citep{2011ApJ...728...88G}:
\begin{align*}
\Lambda &= \ln\left( 1 + g \DMW^{3/7} (\UMW/15)^{4/7}\right), \\
g &= \frac{1+\alpha{}s + s^2}{1+s}, \\
s &= \frac{0.04}{1.5\times{}10^{-3}\,\ln(1+(3\UMW)^{1.7}) + \DMW}, \\
\alpha &= 2.5\frac{\UMW}{1+(\UMW/2)^2}.
\end{align*}

This model has the following properties:

\paragraph{The asymptotic slope:} The ratio of the $\H2$ to the $\HI$ surface density is
\[
R_{\rm mol} = \frac{2\Sigma_g}{\Sigma^\infty_\HI}\frac{1}{2 + \frac{\Sigma^{\infty}_\HI}{2\Sigma_g}}.
\]
Hence, $R_{\rm mol}\sim{}\Sigma_g/\Sigma^\infty_\HI$ for $\Sigma_g\gg{}\Sigma^{\infty}_\HI$ and $R_{\rm mol}\sim{}4(\Sigma_g/\Sigma^\infty_\HI)^2$ for  $\Sigma_g\ll{}\Sigma^{\infty}_\HI$, respectively. Following the same argument as outlined in the beginning of this section we infer that $P_{\rm ext}\propto{}\Sigma_g^{(2+\beta)/2}$. Here, $\beta$ is 1 if the mid-plane density is dominated by the gas component, otherwise $\beta$ is the exponent of the assumed $\Sigma_*\propto{}\Sigma_g^\beta$ scaling. Combining these results we find that in the $\H2$ dominated regime
\[
R_{\rm mol} \propto{} P_{\rm ext}^{2/(2+\beta)},
\]
i.e. $R_{\rm mol} \propto{} P_{\rm ext}^{0.67}$ for gas dominated galaxies, while in the $\HI$ dominated regime
\[
R_{\rm mol} \propto{} P_{\rm ext}^{4/(2+\beta)},
\]
i.e. $R_{\rm mol} \propto{} P_{\rm ext}^{1.33}$ for gas dominated galaxies. These slopes are reasonably close to the slopes $\sim{}0.8$ ($\H2$ dominated regime) and $\sim{}1.2$ ($\HI$ dominated regime) measured just above and below the transition regime by \cite{2008AJ....136.2782L}.

\paragraph{The normalization:} The model reproduces the variation of the normalization of the $\log_{10} P_{\rm ext} - \log_{10} R_{\rm mol}$ relation with dust-to-gas ratio as demonstrated by Fig~\ref{fig:FitTheo}. This implies that the decrease in the normalization with decreasing $\DMW$ is primarily a result of the increase of the $\HI$ saturation limit with increasing dust-to-gas ratio. The model predicts that in the $\H2$ dominated regime, to first order, $\Sigma_\HI^\infty\propto\DMW^{-1}$ and, hence, $R_{\rm mol}\propto{}\DMW{}$ at a fixed $\Sigma_g$. We therefore expect the normalization of the $\log_{10} P_{\rm ext} - \log_{10} R_{\rm mol}$ relation to scale roughly linearly with the dust-to-gas ratio, in agreement with what our numerical simulations predict. In the $\HI$ dominated regime $R_{\rm mol}\propto{}(\Sigma^\infty_\HI)^{-2}$ and we expected a steeper ($\DMW^2$) dependence of the normalization on the dust-to-gas ratio.

Observations that convert $\CO$ emission into $\H2$ surface densities using a constant conversion factor will find an even faster decline in the normalization with dust-to-gas ratio. A scaling of the conversion factor with dust-to-gas ratio of the form $\DMW^{-\gamma}$ implies that the decrease of the normalization with dust-to-gas ratio is enhanced by the additional factor $\DMW^{\gamma}$. This effect can be seen quite clearly in Fig.~\ref{fig:FitTheo}. In fact, if the model outlined in this section is adopted as a baseline, the dust-to-gas ratio dependence of the conversion factor could be inferred from an observational study of how the normalization of the $\log_{10} P_{\rm ext} - \log_{10} R_{\rm mol}$ relation changes with the dust-to-gas ratio (or metallicity) of the galaxy.

\paragraph{Limitations:}
As pointed out, the model predicts the scaling $\propto{}P_{\rm ext}\propto{}\Sigma_g^{(2+\beta)/2}$ in the $\H2$ dominated regime with $\beta\sim{}1$ if the galaxy is gas dominated. Yet, while many of the observations have a slope consistent with this prediction, some have a significantly larger slope of about unity. In addition, the left panel of Fig.~\ref{fig:FitObs} reveals that the slope becomes smaller if the UV field is increased and that it becomes larger as the dust-to-gas ratio is decreased. How do we explain these trends?

Formally, a value $\beta$ close to zero could explain a slope $\sim{}1$ in some observed galaxies. However, using the \cite{2008AJ....136.2782L} data we find that $\beta$ lies in the range $1-2$ for most galaxies in their sample. We would thus expect slopes only in the range $\sim{}0.5-0.7$. Instead, the answer is that in those galaxies the $\HI$ surface density starts to \emph{decrease} with increasing $\Sigma_g$ at large values of the neutral gas surface density. Note that such a decrease is not present in the model, which instead predicts a monotonic increase of $\Sigma_\HI$ with $\Sigma_g$, see (\ref{eq:ModelSigmas}). However, a drop of $\Sigma_\HI$ with increasing $\Sigma_g$ has been observed before, particularly in centers of galaxies \citep{1978ApJ...223..803M, 2002ApJ...569..157W} although its origin remains unknown. The decrease of $\Sigma_\HI$ with $\Sigma_g$ boosts the increase of $R_{\rm mol}$ with $P_{\rm ext}$ and leads to a steeper slope. 

The trend of a decreasing slope with increasing $\UMW$ appears to be driven largely by the scatter in $\Sigma_\HI$ at a given $\Sigma_g$. Specifically, plotting $\Sigma_\HI$ vs $\Sigma_g$ reveals that while $\Sigma_\HI$ does not exceeds its saturation limit, it does frequently lie below it. This downward-only scatter can be as large as ~1 dex at $\Sigma_g\sim{}30$ $M_\odot$ pc$^{-2}$ and gradually decreases with increasing $\Sigma_g$ out to large surface densities ($\Sigma_g\sim{}1000$ $M_\odot$ pc$^{-2}$). This change in the scatter leads to an overall increase in the median $\Sigma_\HI$ with $\Sigma_g$. Consequently, $R_{\rm mol}$ increases less quickly with $P_{\rm ext}$ and, hence, a reduced slope is obtained, see Fig.~\ref{fig:Fit2}.

Finally, the increase in the $\HI$ saturation limit with decreasing dust-to-gas ratio leads to the anti-correlation between slope and $\DMW$. When the dust-to-gas ratio is decreased, a higher gas surface density and, hence, $P_{\rm ext}$ needs to be reached in order to fully convert the gas from $\HI$ to $\H2$. Therefore, a significant fraction of the ISM regions with $\Sigma_\H2$ above the detection threshold contain a large, or even a dominating, contribution of $\HI$ and  $R_{\rm mol}>1$ is often not fulfilled. This is illustrated in Fig.~\ref{fig:Fit2} which shows that cells above the truncation limit can have $\H2$ to $\HI$ ratios as small as 1\%. The figure also shows that $R_{\rm mol}$ has a much steeper\footnote{Empirically it is not too hard to understand what happens at $R_{\rm mol}<1$. The surface densities of the star formation rate scales super-linearly with $\Sigma_g$ near the so-called threshold of the Kennicut-Schmidt relation \citep{1959ApJ...129..243S, 1963ApJ...137..758S, 1998ApJ...498..541K}. In this regime the atomic hydrogen is not yet fully saturated and often $R_{\rm mol}<1$. Since $\Sigma_{\rm SFR}\propto{}\Sigma_\H2$ \citep{2002ApJ...569..157W, 2008AJ....136.2846B}, $\Sigma_\H2$ and, thus, $R_{\rm mol}$ scale super-linearly with $\Sigma_g$ in this regime. Hence, $R_{\rm mol}$ will also scale super-linearly with $P_{\rm ext}$.} dependence on $P_{\rm ext}$ in the $\HI$ dominated regime ($R_{\rm mol}<1$). Hence, including data points that follow this steeper dependence will bias the fit high. A possible solution to avoid this problem is to restrict the fit to data that falls either only in the $\H2$ or in the $\HI$ dominated regime.

\section{Summary and Discussion}
\label{sect:summary}

We studied the relation between mid-plane pressure and the $\Sigma_\H2/\Sigma_\HI$ ratio in a set of hydrodynamical simulations that follow explicitly the formation and destruction of molecular hydrogen in the ISM. We have demonstrated that these simulation predict a $P_{\rm ext}-R_{\rm mol}$ relation that is very similar to the one observed in nearby galaxies. The fact that the relation changes systematically with the dust-to-gas ratio and the strength of the UV radiation field in the ISM indicates that the $P_{\rm ext}-R_{\rm mol}$ relation should not be used as a universal tool to estimate $\H2$ abundances in galaxies with ISM conditions different from those in the Milky Way. In particular, we find that the normalization of the $\log_{10} P_{\rm ext} - \log_{10} R_{\rm mol}$ relation decreases with a decreasing dust-to-gas ratio $\DMW$. The scaling of the normalization is $\propto\DMW$ as long as the ISM regions that are included in fitting the $\log_{10} P_{\rm ext} - \log_{10} R_{\rm mol}$ relation are predominantly molecular and the metallicity dependence of the $\CO$-$\H2$ conversion factor is taken into account. We have proposed a simple model that is based on the numerical study of the $\HI$ to $\H2$ transition that captures many of the basic properties of the $P_{\rm ext}-R_{\rm mol}$ relation.

\cite{2008ApJ...680.1083R} studied the $P_{\rm ext}-R_{\rm mol}$ relation in hydrodynamical simulations of isolated disk galaxies. While their numerical approach differs in detail, they also find a clear correlation between $P_{\rm ext}$ and $R_{\rm mol}$ with a slope $\sim{}0.9$. Although most ISM regions included in their analysis probe the $\HI$ dominated regime, their largest galaxy reaches pressures high enough to study the $P_{\rm ext}-R_{\rm mol}$ relation in molecular hydrogen dominated gas. Interestingly, while our numerical modeling predicts a change in the slope of the relation from $\sim{}1.3$ for $R_{\rm mol}\ll{}1$ to $\sim{}0.7$ for $R_{\rm mol}\gg{}1$, such a change is not immediately apparent in the work by \cite{2008ApJ...680.1083R}. As showed in the preceding section, a slope of $\lesssim{}0.7$ in the $\H2$ dominated regime is a direct consequence of a fixed $\HI$ saturation limit. Hence, the relatively steep slope ($>0.7$) found by \cite{2008ApJ...680.1083R} in the $\H2$ dominated regime might indicate that in their numerical model $\Sigma_\HI$ decreases with increasing $\Sigma_g$ at large gas surface densities.

\cite{1993ApJ...411..170E} studied the origin of the relation between $\H2$ abundance and mid-plane pressure analytically. He found that in the $\HI$ dominated regime $R_{\rm mol}\propto{}f_\H2\propto{}P_{\rm ext}^{2.2}/j$, where $j$ is the local radiation field. While this relation is sometimes used to ``explain'' the $P_{\rm ext}-R_{\rm mol}$ relation, we stress that the $\HI$ dominated regime is not the regime typically studied in observations. Furthermore, $j$, the interstellar radiation field incident on the clouds in the ISM, is likely highly spatially variable and will correlate only to a certain degree with galactic properties smoothed over $\sim{}$kpc scales. This implies that applying this formula properly is non-trivial. For instance, if the $\sim{}$kpc scale UV radiation field is approximately constant a naive application of Elmegreen's formula predicts $R_{\rm mol}\propto{}P_{\rm ext}^{2.2}$. In contrast, the model presented in section \ref{sect:model} predicts $R_{\rm mol}\propto{}P_{\rm ext}^{4/(2+\beta{})}$ for $R_{\rm mol}<1$, which results in a scaling $R_{\rm mol}\propto{}P_{\rm ext}^{1-1.33}$ (for $\beta{}\sim{}1-2$) close to observations \citep{2008AJ....136.2782L}.

The numerical predictions presented in this paper are based on the modeling of a variety of well understood physical processes in the ISM. However, it is possible or even likely that we are missing processes that have an impact on the atomic and molecular hydrogen abundance in galaxies. For instance, while the simulations include certain aspects of stellar feedback (e.g., ionizing and dissociating radiation from massive stars, metal enrichment from supernovae, stellar mass loss), they miss others (e.g., radiation pressure, thermal feedback from supernovae, stellar winds). Also, the resolution of the simulations is too coarse to resolve individual molecular clouds and the detailed dynamics and small scale chemistry that takes place within them. Hence, it will be important to verify the predictions of our numerical models observationally. Important tests include (i) the measurement of the normalization of the $\log_{10} P_{\rm ext} - \log_{10} R_{\rm mol}$ relation in galaxies with low metallicities and presumably low dust-to-gas ratios, (ii) the measurement of the slope of the relation in both the $\HI$ and $\H2$ dominated region, and (iii) to check for correlations of the normalization and slope with the scaling exponent between the stellar surface density and the gas surface density. 

\acknowledgements 

RF thanks Andrey Kravtsov and Adam Leroy for stimulating discussions during the "Galactic Scale Star Formation" conference in Heidelberg 2012. This work was supported in part by the DOE at Fermilab, by the NSF grant AST-0708154, by the NASA grant NNX-09AJ54G, and by the Kavli Institute for Cosmological Physics at the University of Chicago through the NSF grant PHY-0551142 and PHY-1125897 and an endowment from the Kavli Foundation. The simulations used in this work have been performed on the Joint Fermilab - KICP Supercomputing Cluster, supported by grants from Fermilab, Kavli Institute for Cosmological Physics, and the University of Chicago. This work made extensive use of the NASA Astrophysics Data System and {\tt arXiv.org} preprint server.

\bibliographystyle{apj}

\end{document}